\documentclass[reprint,amsmath,amssymb,superscriptaddress,prc,floatfix,nofootinbib]{revtex4-1}
\usepackage{xcolor}
\usepackage{amsmath}
\usepackage{graphicx}
\usepackage{longtable}
\usepackage{bm}
\usepackage[caption=false,subrefformat=parens]{subfig}
\usepackage{hyperref}

\hypersetup{colorlinks=true,breaklinks,linkcolor=red,citecolor=blue}

\begin{document}
  
\title{Nuclear ground states in a consistent implementation of the time-dependent density matrix approach}

\author{Matthew Barton}
\affiliation{Department of Physics, Faculty of Engineering and Physical Sciences, University of Surrey, Guildford, Surrey GU2 7XH, United Kingdom}

\affiliation{Nuclear Theory Group, Faculty of Physics, Warsaw University of Technology, 00-662 Warsaw, Poland}

\author{Paul Stevenson}
\email{p.stevenson@surrey.ac.uk}
\affiliation{Department of Physics, Faculty of Engineering and Physical Sciences, University of Surrey, Guildford, Surrey GU2 7XH, United Kingdom}

\author{Arnau Rios}
\email{a.rios@surrey.ac.uk}
\affiliation{Department of Physics, Faculty of Engineering and Physical Sciences, University of Surrey, Guildford, Surrey GU2 7XH, United Kingdom}
\affiliation{
 Departament de F\'isica Qu\`antica i Astrof\'isica, 
 Institut de Ci\`encies del Cosmos (ICCUB),
Universitat de Barcelona, Mart\'i i Franqu\`es 1, E08028 Barcelona, Spain
}%

\date{\today}

\begin{abstract}
\begin{description} 
\item[Background] Time-dependent techniques in nuclear theory often rely on mean-field or Hartree-Fock descriptions. Beyond mean-field dynamical calculations within the time-dependent density matrix (TDDM) theory have often invoked symmetry restrictions and ignored the connection between the mean-field and the induced interaction. 
\item[Purpose] We study the ground states obtained in a TDDM approach for nuclei from $A=12$ to $A=24$, including examples of even and odd-even nuclei with and without intrinsic deformation. We overcome previous limitations using three-dimensional simulations and employ density-independent Skyrme interactions self-consistently. 
\item[Methods] The correlated ground states are found starting from the Hartree-Fock solution, by adiabatically including the beyond-mean-field terms in real time. 
\item[Results] We find that, within this approach, correlations are responsible for $\approx 4-5 \%$ of the total energy. Radii are generally unaffected by the introduction of beyond mean-field correlations. Large nuclear correlation entropies are associated to large correlation energies. 

By all measures, $^{12}$C is the most correlated isotope in the mass region considered. 
\item[Conclusions] 
Our work is the starting point of a consistent implementation of the TDDM technique for applications into nuclear reactions. Our results indicate that correlation effects in structure are small, but 

beyond-mean-field dynamical simulations could provide new insight into several issues of interest.  

\end{description}
\end{abstract}

\pacs{}

\maketitle

\section{Introduction}

The study of the time evolution of nuclei provides key insights into their structure, their excitation and their associate reactions. Several techniques have been devised to tackle numerically the dynamics of nuclear many-body systems \cite{rin80,Bla86,Simenel2008,Simenel2012}. Traditionally, non-stationay simulations in nuclear physics have been implemented in the time-dependent Hartree-Fock approximation (TDHF) \cite{Engel1975,Bonche1976,Kim97} or, in more modern terms,  the time-dependent density functional approach \cite{Nakatsukasa2016,Bulgac2013,Bulgac2016}. This approach assumes that nucleons move only under an (instantaneous) average potential generated by the other nucleons and consistently takes the Pauli exclusion principle into account \cite{KohnSham65}. Using Skyrme density functionals, simulations are nowadays routinely implemented in unrestricted three-dimensional (3D) geometries and have been used to describe a plethora of different nuclear phenomena \cite{Umar2006,Bulgac2011,Scamps2014,Scamps2015,Stevenson2016,sky3dcode,Iwata2019,Nishikawa2018,Stevenson2020,Godbey2020}.

In the past, there have been attempts to move beyond this mean-field approximation. There are a handful of methods that introduce genuine two-body correlations in the dynamics \cite{nonequilibrium}. These include, amongst others,  the Balian-V{\'e}neroni approach to incorporate particle-number fluctuations \cite{Bal81,Bro08,Sim11} and the Kadanoff-Baym approach for infinite \cite{Danielewicz1984a,Kohler1999} and finite \cite{Mahzoon2017,Lin2020} systems. 
However, the Time-Dependent Density Matrix (TDDM) approach is probably the most widely applied beyond-mean-field method in nuclear physics and, in this sense, it is the most successful to date.

TDDM was first introduced by Cassing and Wang in 1980 \cite{Cassing1980}, and has been applied extensively in nuclear physics by Tohyama  \cite{Tohyama1987,Gong1990,Tohyama1994,Tohyama1998,Tohyama1998b,Tohyama1999,Tohyama2002,Tohyama2002b,Tohyama2013,Tohyama2015,Tohyama2016,Tohyama2019,Tohyama2020} and others \cite{GongPhD,Cassing1990,Cassing1992}. In this context, the TDDM equations are often projected into a moving basis dictated by a TDHF-like equation, plus a time-dependent term that depends on correlations \cite{GongPhD}. Successful implementations with different levels of consistency have also been used to study breakup \cite{Assie2009,AssiePhD} and, recently, fusion reactions in an energy-conserving approach \cite{Wen2018,Wen2019}.

TDDM allows one to go beyond TDHF by truncating the Bogoliubov--Born--Green--Kirkwood--Yvon (BBGKY) hierarchy of quantum mechanical many-body density matrices order by order \cite{nonequilibrium,manyparticletheory}. 
Here, and in the following, we define uncorrelated systems as those where the probability distributions of two particles are independent. Two-body correlations are therefore a measure of the ``lack" of independence of the two probability distributions. This approach has also seen widespread use within other areas of physics, such as condensed matter and quantum optics, where TDDM often goes by the name of reduced density matrix theory \cite{Akbari2012,AkbariPhD,BreuerPetruccione}.

In addition to the many-body truncation, other approximations are typically implemented in nuclear TDDM simulations. Due to computational intensiveness of the calculations, previous implementations of TDDM have worked in spherical or, more recently, axial symmetry \cite{Tohyama2002,Tohyama2016}. Moreover, one often assumes that a decoupling applies to nuclear systems, so that the interaction acting at the mean-field level is different to that acting at the beyond mean-field level \cite{Danielewicz1984a}. The latter is often dubbed the ``residual interaction" and often takes the form of a simplified $\delta-$function \cite{Tohyama2002b,Assie2009,Wen2018}. 
On the one hand, the geometric restrictions preclude the application of TDDM methods to triaxially deformed nuclei and to general dynamical settings between multiple nuclei. On the other, the inconsistent use of mean-field and residual interactions hampers the possibility of discussing systematics in the TDDM expansion. For instance, if one were to find an improvement in the dynamical description when going from TDHF to TDDM, it would be difficult to unambiguously ascribe the improvement to the use of TDDM when the employed Hamiltonian is not the same at all levels. 

In view of these limitations, we have implemented a fully unrestricted 3D implementation of TDDM that uses the Skyrme interaction for both the mean-field and the residual channels \cite{BartonPhD}. Our truncation of the BBGKY hierarchy includes two-body correlations only~\cite{Tohyama2019}. In this paper, we provide details and results for our implementation of this method. Additional information can be found in the PhD thesis of Ref~\cite{BartonPhD}. 

Our physics focus is the generation of correlated ground states within the TDDM approach. We obtain these from the dynamical equations by means of an adiabatic switching-on technique, as explained below. Our calculations extend from light systems, like $^{12}$C, up to exotic nuclei, like $^{24}$O. We do not expect the calculations to provide a good match to experimental data, because the interactions have been fitted at the mean-field level. However, the simulations provided here are informative in terms of the structure, size and mass evolution of two-body correlations within the TDDM formalism with Skyrme forces. Ultimately, our aim is to use the ground states described here to study nuclear dynamics within a fully consistent TDDM approach.

This paper is laid out as follows. Section~\ref{sec:technical} gives a brief outline of the theoretical background and the numerical implementation of our TDDM approach. Some further details are provided in the Appendix. 

In Sec.~\ref{sec:hf}, we discuss nuclear ground states obtained within HF calculations, which are necessary for comparison to the correlated TDDM results provided in Sec.~\ref{sec:correlated}. 

Section~\ref{sec:conclusions} concludes this paper with a summary and short discussion on areas of future research.

\section{Time-dependent density matrix method} 
\label{sec:technical}

\subsection{Formalism}

The BBGKY hierarchy relates the time evolution of an $A-$body density matrix, $\rho_A$, to the Hamiltonian, $\hat H$, and the $(A+1)$ density matrix \cite{Cassing1980}. 
A truncation of the hierarchy is necessary to make the dynamical equations of the density matrix numerically tractable for practical implementations. Depending on the truncation, one finds different coupled differential equations for the evolution of the density matrix that obey conservation laws \cite{Akbari2012,AkbariPhD}. 
 By assuming that $A=3$ body correlations can be cast in terms of $A=2$ and $A=1$ density matrices only, one recovers the most popular implementation of TDDM \cite{Cassing1990,Tohyama2014,*Tohyama2017,Tohyama2019}. 

If we denote generally the space coordinates of a nucleon by $x_i$, the two-body density matrix (with no reference to spin or isospin) is a tensor in four positions, $\rho_2(x_1',x_2'; x_1, x_2)$. In a 3D mesh of $N_x$ points in each direction, this quantity scales like $N_x^{12}$, quickly overcoming present computational capabilities. To avoid this limitation, we solve the TDDM equations in a moving TDHF-like basis \cite{GongPhD}. This has several advantages. First, because part of the dynamics is incorporated in the basis, the TDDM equations are simplified with respect to static basis approaches \cite{Cassing1990}. More importantly, the size of the correlation tensor is dictated by the total number of single-particle orbitals, $N_\text{max}$, and scales with the fourth power of this variable, $N_\text{max}^4$.  In addition, we can use already existing computational capabilities at the TDHF level to evolve the basis states in a fully unrestricted 3D geometry \cite{sky3dcode}. We note, however, that there are instances, particularly in fusion reactions in the merging phase, where a finite basis set may be insufficient to guarantee energy conservation \cite{Wen2018,Wen2019}.

In this approach, the one-body density matrix is expanded into a finite set of HF-like single-particle orbitals that depend on time,
\begin{align}
\rho_1 (x_{1},x_{1}'; t)  &=  \sum_{\alpha \alpha'}^{N_\mathrm{max}} n_{\alpha \alpha'}(t) \psi_{\alpha'}^{*} (x_{1}',t) \psi_{\alpha} (x_{1},t) \, .
\label{eq:rho1}
\end{align} 
In this subsection, we denote by $N_\text{max}$ the total number of such states, including neutrons and protons\footnote{In the numerical implementations discussed below, we also denote the total number of neutron and of proton states, independently, by $N_\text{max}$. The factor of $2$ between the two definition should not cause any confusion.}.
From now on we omit the time variable $t$ as it is clear that all quantities depend on it and all our summations run from the lowest single-particle index up to $N_\text{max}$. 
The single-particle orbitals follow the dynamics dictated by a TDHF-like equation, 
\begin{align}
i \hbar \dfrac{d}{dt} \psi_{\lambda} = \sum_{\alpha} \epsilon_{\alpha \lambda}  \psi_{\alpha} \, ,
\label{eq:spTDHF}
\end{align}
where 
\begin{align}
\epsilon_{\alpha \beta} = t_{\alpha \beta} + \sum_{\gamma \delta} \nu_{\alpha \gamma \beta \delta} n_{ \delta \gamma}.
\end{align}
is the so-called energy matrix. This includes a kinetic contribution, $t_{\alpha \beta}$, and an interaction term. We give more details on the calculation of the matrix elements $\nu_{\alpha \gamma \beta \delta \gamma}$ below. If the energy matrix is diagonal, the elements $\epsilon_{\alpha \alpha}$, are the single-particle energies associated with a given state $\alpha$. 

The matrix $n_{\alpha \alpha'}$ is known as the occupation matrix. When the occupation matrix is diagonal, the diagonal elements correspond to the occupations of the associated single-particle orbitals. The time evolution of $n_{\alpha \alpha'}$ is dictated by the correlation tensor, $C$. The latter corresponds to the correlated part of the two-body density matrix, $C= \mathcal{A}(\rho_1 \rho_1) - \rho_{2}$, where the operator $\mathcal{A}$ antisymmetrizes with respect to exchanges between single indices $x_i$ and $x_j$. The correlation tensor can also be decomposed into a time-dependent single-particle basis,
\begin{widetext}
\begin{align}
C ( x_{1},x_{2};x_{1}',x_{2}' )  &= \sum_{\alpha \beta \alpha' \beta'}^{N_\mathrm{max}} C_{\alpha \beta \alpha' \beta'} \psi_{\alpha'}^{*} (x_{1} ) \psi_{\beta'}^{*} (x_{2} ) \psi_{\alpha} (x_{1}') \psi_{\beta} (x_{2}').
\end{align}
\end{widetext}
Upon making this decomposition, one finds that the evolution of $n_{\alpha \alpha'}$ becomes:
\begin{equation}
i \hbar \dfrac{d n_{\alpha \alpha'}}{dt} = \sum_{\gamma \delta \sigma}^{N_\mathrm{max}} \Big[ 
  \nu_{\alpha \sigma \gamma \delta} C_{\gamma \delta \alpha' \sigma} 
- C_{\alpha \delta \gamma \sigma} \nu_{\gamma \sigma \alpha' \delta} \Big] \, .
\label{eq:occmat}
\end{equation}
As opposed to a static basis projection, the right-hand side of this equations does not have any Hartree-Fock (HF) term \cite{Cassing1990,GongPhD}. 
As one can clearly see, when correlations are not active ($C=0$), the occupation probabilities become static. Further, in a pure mean-field picture without pairing correlations, one can prove that these  occupation probabilities are either $1$ or $0$ depending on whether the orbital is below or above the Fermi surface, respectively. 

In contrast, when correlations are active, one expects that the occupation numbers take values between $0$ and $1$, to abide with the Pauli principle and their probabilistic nature. The truncation in the TDDM equations does not always mathematically guarantee that this is the case, as numerically corroborated in early nuclear physics applications \cite{Schmitt1990,Gherega1993} and more recent strongly correlated electronic simulations~\cite{Akbari2012}. We have observed this anomalous behaviour in a handful of simulations, but it is difficult to discriminate their origin that could partially be due to numerical issues. 

When the hierarchy is truncated at some level, the evolution of the correlation tensor $C$ is dictated by an equation which depends on occupation numbers; interaction matrix elements; and the correlation tensor itself. We work under the assumption that genuine three-body correlations are negligible \cite{GongPhD,Akbari2012,Tohyama2020}. In other words, the three-body density matrix is a properly antisymmetrized product of one-body and two-body density matrices only, but does not include any genuine $C_3$ terms. Under this approximation, the equation of motion for the correlation tensor is \cite{BartonPhD} :
\begin{widetext}
\begin{align} \label{corten}
i \hbar \dfrac{d C_{\alpha \beta \alpha'  \beta'  }}{dt}  &
= \ \dfrac{1}{2} \sum_{\lambda  \mu  } \nu_{\beta \alpha \lambda  \mu  } ( n_{\lambda \beta'  } n_{\mu  \alpha'  } - n_{\lambda  \alpha'  } n_{\mu  \beta'  } + C_{\mu  \lambda \alpha'  \beta'  }) 
+ \dfrac{1}{2} \sum_{\lambda  \mu} \nu_{\lambda  \mu\alpha'  \beta'  } (n_{\beta \lambda } n_{\alpha \mu} - n_{\beta \mu} n_{\alpha \lambda} + C_{\alpha \beta \mu \lambda}) \nonumber \\
&- \dfrac{1}{2} \sum_{\delta \lambda \mu  } \nu_{\beta \delta  \lambda \mu  } \Bigg[ n_{\lambda \beta'  } (n_{\mu  \alpha'  } n_{\alpha \delta } - n_{\alpha \alpha'  } n_{\mu  \delta } +  C_{\mu  \alpha \alpha'  \delta  } ) - n_{\lambda \alpha'  } C_{\mu \alpha \beta'  \delta } - n_{\mu  \beta'  } C_{\lambda \alpha \alpha'  \delta } + n_{\mu  \alpha'  } C_{\lambda \alpha \beta'  \delta } + n_{\alpha \delta } C_{\lambda \mu  \beta'  \alpha'  } \Bigg] \nonumber \\
&+ \dfrac{1}{2} \sum_{\delta \lambda \mu  } \nu_{\delta \lambda \beta'  \mu  } \Bigg[  n_{\beta \lambda} (n_{\mu  \delta} n_{\alpha \alpha'  } - n_{\alpha \delta} n_{\mu  \alpha'  } - C_{\alpha \mu \delta \alpha'  } ) + n_{\beta \delta} C_{\alpha \mu  \lambda \alpha'  } - n_{\alpha \delta} C_{\beta \mu \lambda \alpha'  } + n_{\alpha \lambda} C_{\beta \mu \delta \alpha'  }  + n_{\mu  \alpha'  } C_{\beta \alpha \delta \lambda} \Bigg] \nonumber \\
&+ \dfrac{1}{2} \sum_{\delta \lambda \mu  } \nu_{\alpha \delta \lambda \mu  } \Bigg[ n_{\lambda \beta'  } (n_{\mu  \alpha'  } n_{\beta \delta} - n_{\beta \alpha'  } n_{\mu \delta} + C_{\mu   \beta \alpha' \delta} ) - n_{\lambda \alpha'  } C_{\mu   \beta \beta'  \delta} - n_{\mu  \beta'  } C_{\lambda  \beta \alpha' \delta} 
+ n_{\mu  \alpha'  } C_{\lambda \beta \beta'  n} + n_{\beta n} C_{\lambda \mu  \beta'  \alpha'  }  \Bigg] \nonumber \\
&-  \dfrac{1}{2} \sum_{\delta \lambda \lambda } \nu_{\delta \lambda \alpha'  \lambda } \Bigg[ n_{\beta \lambda} (n_{\lambda \delta} n_{\alpha \beta'  } - n_{\alpha \delta} n_{\lambda \beta'  } - C_{\alpha \lambda \delta \beta'  } )  + n_{\beta \delta} C_{\alpha \lambda \lambda\beta'  } - n_{\alpha \delta } C_{\beta \lambda \lambda \beta'  }
+ n_{\alpha \lambda} C_{\beta \lambda \delta\beta'  } + n_{\lambda \beta'  } C_{\beta \alpha \delta \lambda} \Bigg] \, . 
\end{align}
\end{widetext}
 We note that this equation uses antisymmetrized matrix elements [see Eq.~(\ref{eq:matels}) below]  unlike other implementations \cite{GongPhD,Cassing1990}. 

 A brute force implementation of the previous equations would scale as $N_\text{max}^{7}$. We exploited the symmetries of the correlation tensor elements to reduce this computational cost  \cite{BartonPhD}. 

 Also, certain matrix elements are zero based on isospin conservation arguments.

The two-body density matrix (or, equivalently, the correlation tensor $C$) provides direct access to the total energy of the system, which is customarily split into two contributions, $E=E_\text{MF}+E_\text{c}$. The mean-field term is expressed in terms of occupation matrix elements only and is already active at the HF level,
\begin{align}
E_\text{MF} = \sum_{\alpha \beta} t_{\alpha \beta} n_{\beta \alpha} 
+ \frac{1}{2} \sum_{\alpha \beta \gamma \delta} \nu_{\alpha \beta \gamma \delta} n_{\gamma 
\alpha} n_{\delta \beta} \, .
\label{eq:emf}
\end{align}
The correlation energy term, in contrast, is directly proportional to $C$,
\begin{align}
E_\text{c} = \dfrac{1}{4} \sum_{\alpha \beta \gamma \delta} \nu_{\alpha \beta \gamma \delta} C_{\delta \gamma \beta \alpha} \, ,
\label{eq:ecor}
\end{align}
and is only active in beyond mean-field calculations. The correlation energy can therefore be used as a metric to quantify correlations in the TDDM approach. 
The TDDM approach based on Eqs.~(\ref{eq:spTDHF}), (\ref{eq:occmat}) and~(\ref{corten})  conserves the number of particles, the total momentum and the total energy \cite{GongPhD}.

\subsection{Interaction matrix elements}

In the past, the implementation of the TDDM approach has often relied on approximations. An often-used assumption is that the beyond mean field interaction (the ``residual interaction") is a delta function multiplied by a constant. This reduces substantially the computational cost required to calculate the matrix elements, $\nu_{\alpha \beta \alpha' \beta'}$. This approximation however ignores the self-consistency between the mean-field interaction and residual interaction which, from a first-principles perspective, should be based on the same Hamiltonian.
In this work, we instead use the Skyrme interaction,
\begin{align}
V({\bf r}) &= 
t_0 (1 + x_0 P_s) \delta({\bf r}) + \frac {1}{6} t_3 (1 + x_3 P_{s}) \rho^{\alpha}({\bf R}) \delta({\bf r}) \nonumber \\
&+ \frac {1}{2} t_1 ( 1 + x_1 P_s) 
\left[ { \bf k'}^2 \delta({\bf r}) + \delta({\bf r}){\bf k}^2 \right]
\nonumber \\
&+ t_2 ( 1 + x_2 P_s) \left[ {\bf k} \cdot \delta({\bf r}) {\bf k} \right] \nonumber \\
&+ i  W_0 ({\bf \sigma}_1 + {\bf \sigma}_2) \cdot \left [ {\bf k}' \times \delta({\bf r}) {\bf k} \right ]\, , 
\label{eq:skyrme}
\end{align}
to model the nucleon-nucleon interaction, both at the mean-field and the residual interaction level. 
In this equation, $\bf r = {\bf r}_1 -{\bf r}_2$ represents the relative distance between two nucleons at positions ${\bf r}_1$ and ${\bf r}_1$; 
${\bf R}= ({\bf r}_1 + {\bf r}_2)/2$, 
${\bf k}= ({\bf \nabla}_1 -{\bf \nabla}_2)/2 i $ the relative momentum
acting on the right and ${\bf k}'$ its conjugate acting on the left. 
$P_s= (1 + \bf{\sigma}_1 \cdot \bf{\sigma}_2)/2$ is the spin exchange
operator. The last term, proportional to $W_0$, corresponds to the zero-range spin-orbit term. 

The use of interactions with density-dependent terms in beyond mean-field implementations may be problematic. These do not constitute true forces and hence must be treated with care in many-body approaches to avoid pathologies \cite{BartonPhD}. We therefore preclude from using standard Skyrme parametrizations, but employ 
two parametrizations of this force, SV \cite{Beiner1975} and SHZ2 \cite{Satula2012}, that do not have a density dependent term. In other words,  $t_3=0$ in the notation of the original Skyrme force \cite{originalskyrme2,sphericalskyrme}. 
We note that SHZ2 is in fact a slight refit of SV, with very similat $t_i$ parameters and a very small $x_0$ term~\cite{Satula2012}. 
These two different fits therefore allow us to minimally explore the parameterization-dependence of our results. 

The matrix elements of the interaction need to be projected into the single-particle orbitals so they can be used in Eqs.~(\ref{eq:occmat}) and (\ref{corten}). This is achieved by means of a double 3D integral
\begin{align}
\nu_{\alpha \beta \alpha' \beta'} = 
\int d {\bf x_1} \, d {\bf x_2}  \, & \psi^*_\alpha({\bf x_1}) \psi^*_\beta({\bf x_2})	
V ( {\bf x_2} - {\bf x_1} ) \times \nonumber \\
& \left[ \psi_{\alpha'}({\bf x_1}) \psi_{\beta'}({\bf x_2}) -  \psi_{\beta'}({\bf x_1}) \psi_{\alpha'}({\bf x_2}) \right] \, .
\label{eq:matels}
\end{align}
However, because of the zero-range nature of the Skyrme force, these integrals simplify substantially \cite{BartonPhD}. 
The calculation of all elements of $\nu_{\alpha \beta \alpha' \beta'}$ scales as $N_{x}^{3} N_{max}^{4}$. This quantity is calculated 4 times at each time-step, which becomes a numerical bottleneck for very fine grids or large boxes, and for heavy systems.
We note that these matrix elements are antisymmetrized from the outset.

\subsection{Adiabatic switching and asymptotic convergence of correlation energies}
\label{asy}

We obtain nuclear ground states employing an adiabatic real time switching technique that continuously transitions from the mean-field to the correlated ground state \cite{BartonPhD}. This is achieved by multiplying the matrix elements of the interaction that appear in Eqs.~(\ref{eq:occmat}) and ~(\ref{corten})  by a 

factor, $\gamma(t)$, that slowly goes from $0$ to $1$. 

We use a Gaussian factor,
\begin{equation}
\gamma (t) = 1 - e^{-\frac{t^{2}}{\tau_{2}}},
\label{gammat}
\end{equation}
which has finite derivatives at $t=0$ and $t \gg 1$. 
We work under the assumption that the switch-on procedure allows the Gell--Mann Low Theorem \cite{GellMannLow1951} to be applied. In other words, if the residual interaction is switched on slowly enough, the final state should become an eigenstate of the TDDM approach. 
Numerical tests indicate that the value $\tau_{2}=32000$ fm$^{2} c^{-2}$ is sufficient to guarantee a converged correlated ground state. 
This correspond to physical changes on a timescale of $t = \sqrt{\tau_2} \approx 180$ fm c$^{-1}$. 

We indicate that a test run with $A=4$ and $\tau_{2}=64000$ fm$^{2} c^{-1}$ provided no significant differences in terms of asymptotic energies.
The asymptotic energy values provided below are obtained either from a converged final result, or from fits of the different energy components assuming a time dependence proportional to $\gamma^2(t)$. 
More details of this procedure can be found in Ref.~\cite{BartonPhD}.

\subsection{Numerical details and bottlenecks}

In unrestricted 3D TDHF simulations, one typically works with as many single-particle orbitals as nucleons in the system, $N_\text{max}=A$ \cite{sky3dcode}.
As nucleons are allowed to scatter off each other in TDDM, the number of single-particle orbitals must necessarily be larger than the number of nucleons, $N_\text{max}>A$. Our TDDM simulations are projected into a TDHF-like basis with an equal maximum number of neutron and proton states, $N_\text{max}$. 
In the following, we provide results for $N_\text{max}=14$ and $20$ to explore what in \emph{ab initio} terms is typically called the ``model-space" dependence of our results. In a shell model language, $N_\text{max}=20$ corresponds to a model space spanning the full $sd$ shell.

There are two major numerical bottlenecks in our approach, that affect  the ability to propagate over time and restrict the size of nuclei that can be tackled. First, simulations are expensive in terms of memory requirements, since the correlation tensor $C$ and the interaction matrix elements both scale like $N_\text{max}^4$ in number of elements. Large amounts of memory are required to store these tensors.
Second, the calculations of both $C_{\alpha' \beta' \alpha \beta}$ and $\nu_{\alpha' \beta' \alpha \beta}$ are time-consuming. As reported before, these scale as $N_\text{max}^{7}$ and $N_\text{max}^{4} N_{x}^{3}$, respectively. 
Generally speaking, for a small model space ($N_\text{max} < N_{x}$), the calculation of the interaction matrix elements takes most of the computational time. For larger model spaces, it is the calculation of $C_{\alpha' \beta' \alpha \beta}$ that dominates the computational cost. We note that parallelization helps in computing some of these tensors at each time step. 

All calculations presented here were performed on a Cartesian 3D grid
with spacings $\Delta x=\Delta y=\Delta z=1$ fm from $-9.5$ to $9.5$ fm, with $N_x=20$. For the relatively light nuclei in consideration here, we operate in a regime where $N_x \approx N_\text{max}$. 
We note that for $^4$He, a smaller grid spacing of $0.5$ fm was tested for both HF and TDDM ground states. The increase in resolution had a negligible impact on any of the computed ground state properties.

As for the computational expense of time propagation, the matrix elements of $C$ and $\nu$ are computed at each time-step, which makes dynamical simulations slow. The three differential equations for the evolution of the single particle orbitals [Eq.~(\ref{eq:spTDHF})], occupations [Eq.~(\ref{eq:occmat})] and correlation tensor elements [Eq.~(\ref{corten})] are solved using a 4 point Runge-Kutta method. We note that the traditional midpoint method commonly used in TDHF \cite{sky3dcode,sky3d11} provided unstable results in the TDDM implementation (unless a very small time step was used). 
In all calculations performed in this work, a value of $dt=0.2$ fm c$^{-1}$ was used. 

Further details about the time stepping procedure are provided in the Appendix.

\section{Mean-field ground states}
\label{sec:hf}

\begin{table}
\centering
  \begin{tabular}{ | l | l | l | l | l | l | l |    }
    \hline
     \multicolumn{7}{|p{5.6cm}|}{SV}  \\ \hline
    & \multicolumn{2}{p{2.1cm}|}{Charge radius [fm]}  & \multicolumn{2}{p{2.1cm}|}{B.E. [MeV] } & \multicolumn{2}{p{2.1cm}|}{Point radius [fm]} \\ \hline
		    Nucleus & HF & Exp & HF & Exp & Proton & Neutron \\
    \hline
        $^{12}$C & 2.724 & 2.4702 & 69.432 & 92.160 & 2.604  & 2.587   \\
    \hline
   		$^{16}$O & 2.765 & 2.6991 & 113.536 & 127.616 & 2.647 & 2.629 \\
    \hline
        $^{20}$Ne & 3.058 & 3.0055  & 138.860 & 160.64 & 2.951 & 2.927 \\
    \hline
        $^{21}$Ne & 3.046 & 2.9695 & 146.79 &  167.391 & 2.939 & 2.991  \\
    \hline
        $^{21}$Na & 3.129 & 3.0136 & 142.926  &  163.065 & 3.025 & 2.919 \\
    \hline
        $^{22}$Na & 3.109 & 2.9852 & 153.626 & 174.130 & 3.004 & 2.978  \\
		\hline
        $^{24}$O & 2.800 & N/A & 144.552 & 168.96 & 2.683 & 3.433 \\
		\hline
    \multicolumn{7}{|p{5.6cm}|}{SHZ2}  \\ \hline
    & \multicolumn{2}{p{2.1cm}|}{Charge radius [fm]} & \multicolumn{2}{p{2.1cm}|}{B.E. [MeV] } & \multicolumn{2}{p{2.1cm}|}{Point radius [fm]} \\ \hline
		    Nucleus & HF & Exp & HF & Exp  & Proton & Neutron
    \\ \hline
       		$^{16}$O & 2.762 & 2.6991 & 113.648 & 127.616 & 2.644 & 2.624  \\
    \hline
       		$^{20}$Ne & 3.054 & 3.0055  & 139.140 & 160.64 & 2.947 & 2.919 \\
    \hline
  \end{tabular}
\caption{Charge radii (columns 2-3) and binding energies (columns 4-5) obtained from HF calculations using the SV (top values) and SHZ2 (bottom values) Skyrme forces, alongside experimental values.  Proton and neutron point radii are reported in columns 6-7.
Different nuclei are listed in each row. Experimental data taken from Refs.~\cite{chartofnuclides,chargeradiuso16}.
}
\label{table:tddmhf}
\end{table}

We start the discussion of results by providing some of the bulk properties of the HF ground states with the SV \cite{Beiner1975} and SHZ2 \cite{Satula2012} Skyrme interactions. These results act as a baseline and  allow us to quantify the importance of the correlations induced by the TDDM approach. The ground states are obtained using Sky3D \cite{sky3dcode} and their properties are summarised in Table \ref{table:tddmhf}. We investigate a wide range of nuclei from $A=12$ to $A=24$.  The top part of Table \ref{table:tddmhf} shows results for the SV force, whereas the bottom part shows SHZ2 results for $^{16}$O, $^{20}$Ne and $^{24}$O. We provide experimental data where known.

Our final aim is not so much to produce a set of reliable ground states to compare to experiments, but rather to develop an understanding of the size and structure of the correlations induced by the TDDM approach.

The charge radii provided in column 2 are generally overestimated with respect to the experimental results (column 3) by about $3 \%$ on average. 
The HF binding energies per nucleon are provided in column 4 of Table \ref{table:tddmhf}. The theoretical results underestimate the experimental ones (column 5) by about $12 \%$ on average. We can put forward some explanations for this discrepancy. First, we note that our results do not include a center of mass correction, which will be relevant for the energetics of the lightest isotopes. In fact, the binding energies are somewhat closer to experimental results as $A$ increases, suggesting this is the case. Second, these effective interactions were fitted to the ground-state properties of spherical system from $A=16$ to $208$, with very many more heavy systems than light isotopes in the fitting protocol. This naturally biases the parametrizations towards heavier nuclei. Finally, the HF approximation is expected to work better for heavier than for light systems on general grounds.  

Overall, however, the HF simulations produce reasonable values of the energy. The mass dependence of the simulations follows reasonably the energy and radius data. We also stress that there are relatively small differences between the results obtained with the two Skyrme interactions.
The charge radii (energies) obtained with SHZ2 are negligibly smaller (larger) than those of SV, in agreement with the fact that this force has a slightly larger saturation density. The relative differences are of the order $\approx 0.1-0.2\%$. We stress again that SV and SHZ2 are very similar parametrizations~\cite{Satula2012}, and
therefore we expect that the correlated TDDM calculations will also show a relatively insignificant parametrization dependence.

\section{Correlated Ground States}
\label{sec:correlated}

We now discuss the results obtained for the correlated TDDM eigenstates. We aim at providing as much of a systematic discussion as possible by focusing on binding energies and radii. We discuss the results isotope by isotope, in order to provide a more detailed explanation and a clearer characterisation of the role of correlations in each of these systems. 

\subsection{\texorpdfstring{$^{12}$C}{12C}}

\begin{table}[tb!] 
\centering
  \begin{tabular}{ | l | l | l |  }
    \hline
    \multicolumn{3}{|p{4cm}|}{$^{12}$C SV}  \\ \hline    
    $N_\text{max}$          & $14$ & $20$  \\ \hline
    $E_\text{c}$ [MeV]     & $-5.5$  & $-8.1$ 
    \\ \hline
    $E_\text{MF}$ [MeV] & $-65.7$ & $-64.0$ 
    \\ \hline
    $E$ [MeV]                 &  $-71.2$ & $-72.1$ 
    \\ \hline
    $E_\text{c}/E$ [\%]  & $7.72$ & $11.23$ 
    \\ \hline
    Proton rms [fm]   & $2.644$ & $2.72 \pm 0.01$ 
    \\ \hline
    Neutron rms [fm]  & $2.627$ & $2.67 \pm 0.01 $ 
    \\ \hline
  \end{tabular}
\caption{Energies (rows 3-5) and radii (rows 6-7) of the $^{12}$C ground states for different values of $N_\text{max}$ (columns 2 and 3). Results are provided for the SV interaction}
\label{table:12c}
\end{table}

The ground state structure of $^{12}$C is of considerable interest for a variety of reasons. In particular, $^{12}$C is relevant because of its possible cluster structure, in which individual nucleons may be correlated with others in a way that cannot be easily captured in a mean-field description \cite{Freer2018}. It is conceivable that the correlations induced by TDDM can capture some of the clusterization mechanisms and provide significantly different ground states. 

\begin{figure}
\centering
  \includegraphics[width=\linewidth]{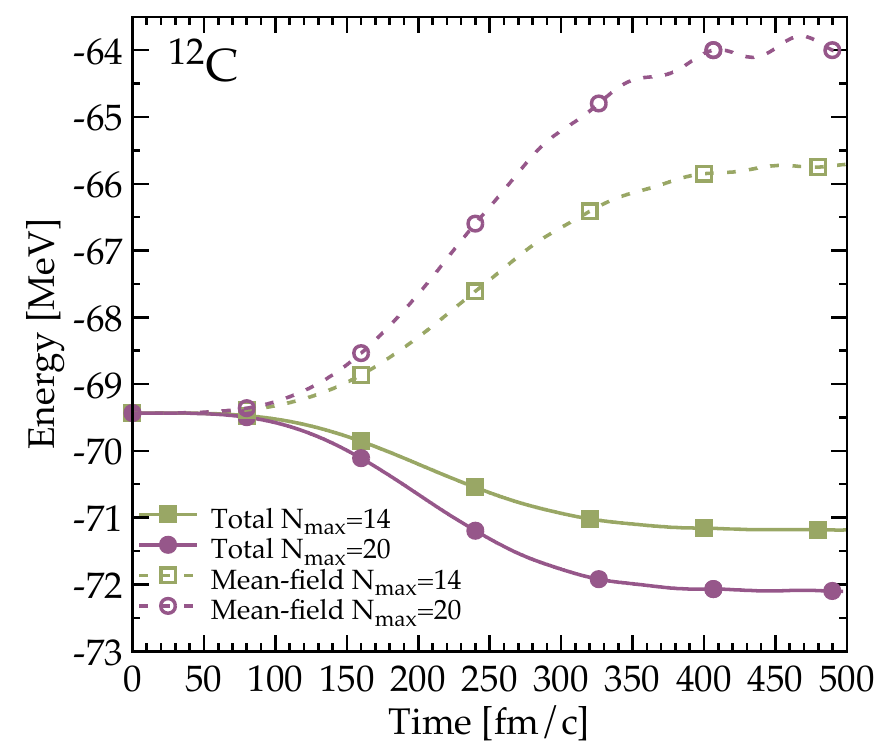}
\caption{
Time evolution of the mean-field (open symbols) and the total energy (filled symbols) of $^{12}$C from a HF to a correlated ground state. Data are provided for $N_\text{max}=14$ (squares) and $N_\text{max}=20$ (circles).
}
\label{fig4:sfig2}
\end{figure}

We summarise the TDDM results for the structure of $^{12}$C in Table \ref{table:12c}. These results are obtained with the SV parametrization. The uncorrelated, HF ground state has a total energy of $E=-69.4$ MeV. 
This is the starting point of the time evolution displayed in Fig.~\ref{fig4:sfig2}, which shows the time evolution of the total (filled symbols) and correlation (empty symbols) energies as a function of time as correlations are switched on. 
Squares (circles) show the results for $N_\text{max}=14$ ($20$). 
As correlations are introduced in the system, the energy  changes. 
The total energy becomes more attractive, whereas the mean-field contribution yields more repulsive results. The total energy drops to $\approx -71$ MeV. In contrast, the mean-field component increases by about $4$ to $5$ MeV. 
Importantly, the final energy is not completely stationary after the evolution finishes at $t=500$ fm/c. 

In relative terms, the correlation energy shown in \ref{table:12c} as a percentage of the total energy is between $7$ and $11 \%$. We anticipate that this is over twice that of any of the other nuclei discussed in the following, which we take as an indication of the importance of correlations for this specific isotope. It would be interesting to find quantitative measures of clustering in these simulations, in line with what has been achieved at the mean-field level~\cite{Ebran2012}.

Columns 2 and 3 of Table \ref{table:12c} show results for both $N_\text{max}=14$ and $20$, respectively. In a traditional shell model picture, the former would include the ground state $1s_{1/2}$ and $1p_{3/2}$ configurations of $^{12}$C as well as the $1p_{1/2}$ and $1d_{3/2}$ levels. 
The larger model space $N_\text{max}=20$ completely fills in the $sd$ shell. We note that, in going from the $N_\text{max}=14$ to $N_\text{max}=20$ configuration, the system gains about $2.5$ MeV of correlation energy, but the nucleus is bound by only $1$ additional MeV. 

As expected, a larger $N_{max}$ corresponds to a larger correlation energy, since the interaction does not have a natural cut-off, and those levels nearest the Fermi energy can be scattered into most available levels. 

The oscillations in energies found at long times in Fig.~\ref{fig4:sfig2}  are evidence of the fact that the system is evolving into the correlated eigenstate too quickly. Turning on the residual interaction more slowly by increasing $\tau_{2}$ in Eq.~(\ref{gammat}) may remedy the oscillations at the end of the calculation, at increased computational cost.  Note, however, that the oscillations in the mean field energy are compensated by anti-phase oscillations in the correlations energy (not shown), giving an overall smooth total energy as a function of time. 

These oscillation are also reflected in the rms radii, which oscillates with a typical size of order $0.01$ fm for the $N_\text{max}=20$ simulation. This uncertainty for radii is reported in the bottom rows of Table~\ref{table:12c}.
Comparing the rms radii to the HF values reported in \ref{table:tddmhf} and the two $N_\text{max}$ values with one another, 

we find that the collisions allow nucleons to scatter further from the nucleus. We note that the proton rms radius increases by about $0.1$ fm, wheres the neutron radius remains relatively constant when increasing the model space. 

\begin{figure}[tb]
\centering
\includegraphics[width=\linewidth]{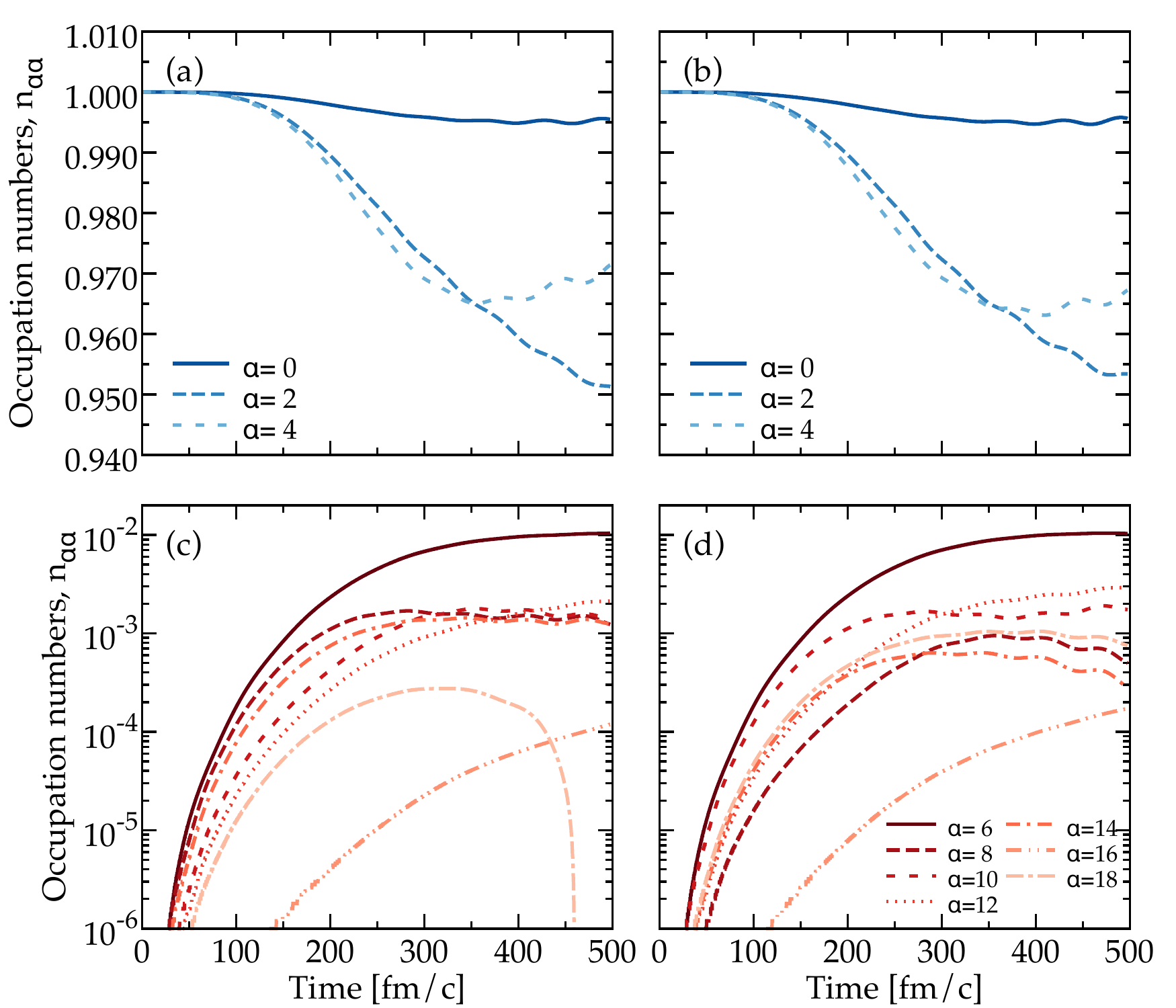}
\caption{Occupation numbers of (a) neutron and (b) proton hole states as a function of time for the adiabatic switching of $^{12}$C with $N_\text{max}=20$. Panels (c) and (d) give the corresponding occupations of particle states in a logarithmic scale. States that are degenerate in spin are not shown for simplicity. The index $\alpha$ denotes energy levels in increasing order.}
  \label{fig:n_C12}
\end{figure}

We can further characterise the correlations in the system by looking at the occupation numbers. 
The time evolution of the neutron and proton diagonal occupation levels, $n_{\alpha \alpha}$, for $^{12}$C is shown in Fig.~\ref{fig:n_C12}.  
The results are shown for the $N_{max}=20$ simulation. Left (right) panels correspond to neutron (proton) states. 
Top panels display the $6$ hole states for both species. Within TDHF, the protons and neutrons completely fill the 1s$_{\frac{1}{2}}$ and 1p$_{\frac{3}{2}}$ sub-shells. 
Bottom panels display particle states instead. 
We find that the mean-field picture is still mostly relevant for the correlated eigenstate in $^{12}$C. 
Hole state occupations reach a value of about $95 \%$. Here, there are clear differences between the more bound $1s_{1/2}$ states, which remain populated to a $99.5\%$ level, and the $1p_{3/2}$ substates, that are substantially more depleted. 
The occupations of particle states are of order $10^{-2}$ or lower. Particle states closer to the Fermi surface, with smaller values of $\alpha$, are more occupied than states further away. 

We note that some states, like the hole $\alpha=0$ ($1s_{1/2}$) and particle $\alpha=6$ ($1p_{1/2}$) are clearly well converged, in the sense that they reach a constant occupation number at large times in the adiabatic switching. Others, in contrast, are still evolving at the end of the simulation. This is the case of the hole $\alpha=2$ state (one of the two $1p_{3/2}$ states shown in short-dashed lines), but also of the $\alpha=18$ neutron state (double-dashed-dotted line in panel c) which has a very low occupation that turns negative just before the end of the evolution. We note that states with large values of $\alpha$ are unbound (eg such that $\epsilon_{\alpha \alpha}>0$, see next paragraph) and hence may be substantially affected by box discretization issues.

\begin{figure}[t]
\centering
\includegraphics[width=\linewidth]{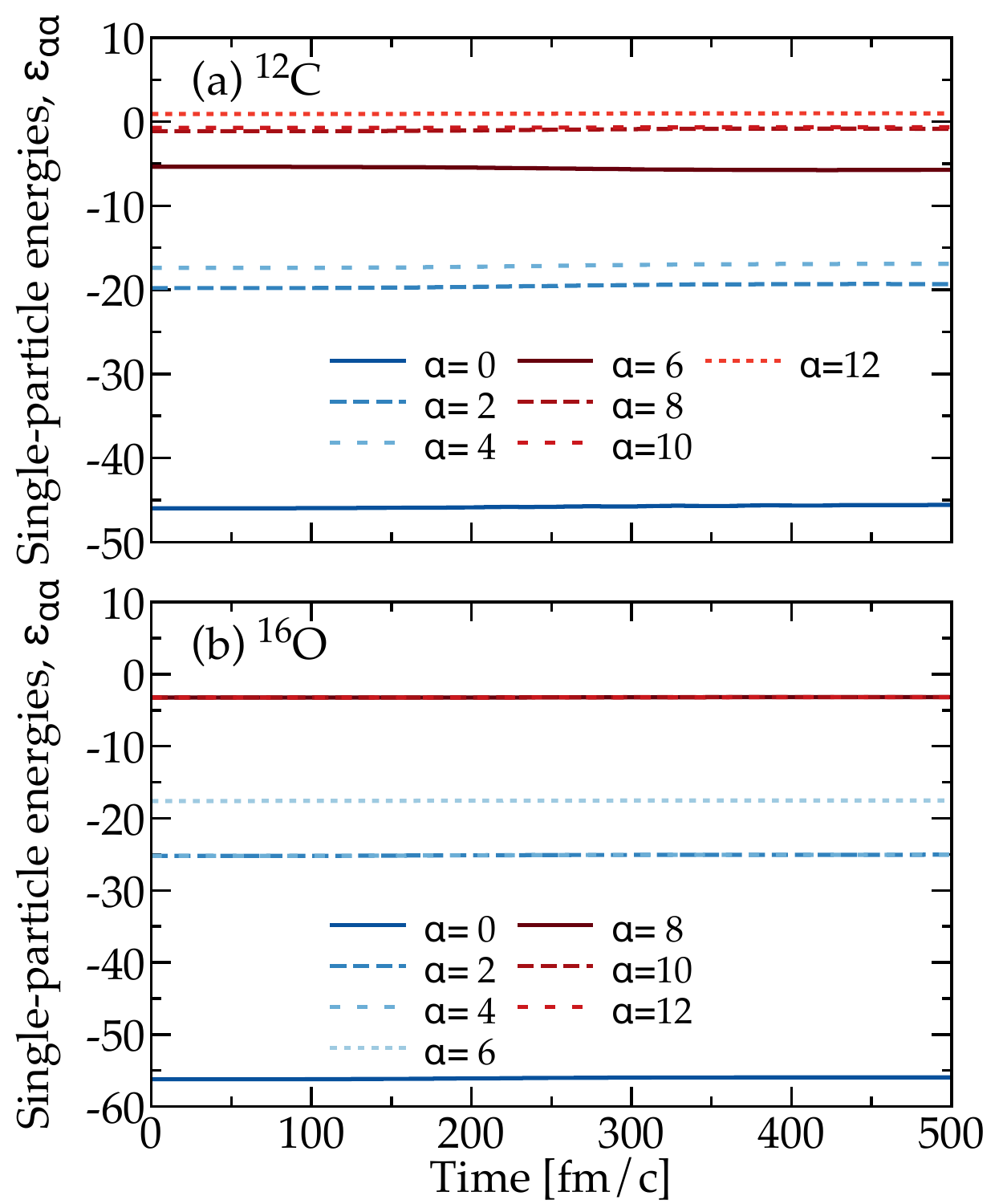}
\caption{Evolution of the single-particle energies $\epsilon_{\alpha \alpha}$ with time for (a) $^{12}$C and (b) $^{16}$O. Particle and hole states are shown in different colors and linestyles. These results are obtained with the SV interaction with $N_\text{max}=14$.
}
  \label{fig:spes}
\end{figure}

As discussed above, $^{12}$C has the largest relative correlation energy of the nuclei we discuss in the following. This may be surprising in the context of Fig.~\ref{fig:n_C12}, which suggests a relatively small redistribution of single-particle strength that could be interpreted as having little impact in the nuclear structure (although, as we shall see later, the changes are not insignificant). 
In addition, the diagonal elements of the single-particle energies themselves do not change much. For $^{12}$C, the time evolution of the $\epsilon_{\alpha \alpha}$ elements are shown in panel (a) of Fig.~\ref{fig:spes} for the $N_\text{max}=14$ simulations. The changes in these single-particle energies are imperceptible in the scale of the graph, in line with previous TDDM implementations \cite{AssiePhD}. To be quantitative, the maximum change across the $500$ fm/c of the simulation, for the most bound $\alpha=0$ ($1s_{1/2}$) state, is less than $0.3$ MeV.

\subsection{\texorpdfstring{$^{16}$O}{16O}}

\begin{figure}[t]
  \centering
  \includegraphics[width=\linewidth]{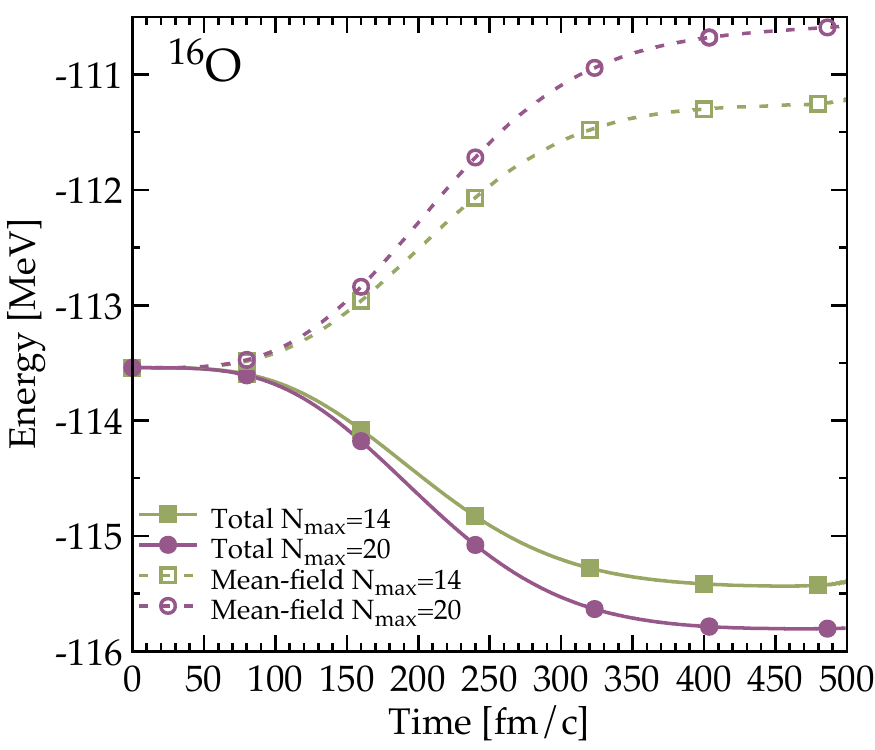}
\caption{Same as Fig.~\ref{fig4:sfig2} for $^{16}$O.}
  \label{fig4:sfig3}
\end{figure}

$^{16}$O is a benchmark nucleus, as a light doubly-magic system which is open to calculation by many beyond mean-field methods. As with $^{12}$C, it is also an $n\alpha$ system, where clustering correlations may play a substantial role. 
A graph of the mean field and total energy of $^{16}$O with the SV Skyrme force parameterisation as it evolves from the HF eigenstate to the correlated eigenstate, for various $N_\text{max}$, is shown in Fig.~\ref{fig4:sfig3}. Simulations start in the HF ground-state at around $\approx -114.5$ MeV. At the end of the adiabatic switching, the total energy is in the range $-115.4-115.9$ MeV (see results in Table~\ref{table:16o}). The correlation energy is about $4$ ($5$) MeV for the $N_\text{max}=14$ ($20$) simulation, whereas the mean-field energy becomes about $\approx2.5$ MeV more repulsive than in the HF case. Overall, the correlation energy contributes about $3.5-4.5 \%$ to the total energy - far less than in the case of $^{12}$C. We also note that no oscillations appear in the total energy or its components in the large-time limit. This may indicate that the transition to the TDDM is somehow ``easier" in this less correlated nucleus.
\begin{table}[b]
\centering
  \begin{tabular}{ | l | l | l | l | l |  }
    \hline
    & \multicolumn{2}{|p{2.1cm}|}{$^{16}$O SV} & \multicolumn{2}{|p{2.1cm}|}{$^{16}$O SHZ2}  \\ \hline    
    $N_\text{max}$          & $14$ & $20$ &  $14$ & $20$  \\ \hline
    $E_\text{c}$ [MeV]     & $-4.1$  & $-5.2$  & $-4.1$ &  $-5.1$  
    \\ \hline
    $E_\text{MF}$ [MeV] & $-111.3$ & $-110.6$ & $-111.4$ & $-110.8$  
    \\ \hline
    $E$ [MeV]                 &  $-115.4$ & $-115.8$ & $-115.5$ & $-115.9$ 
    \\ \hline
    $E_\text{c}/E$ [\%]  & $3.55$ & $4.5$ &  $3.55$ & $4.4$
    \\ \hline
   Proton  & $2.660$ & N/A & $2.657$ & N/A  \\ 
     rms [fm]  &  &    & &   \\ \hline 
   Neutron& $2.640$ & N/A &  $2.636$ & N/A   \\ 
    rms [fm]& &  & &    \\ \hline 
  \end{tabular}
\caption{The same as Table~\ref{table:12c} for $^{16}$O. Results for the SHZ2 parametrization are also provided in columns 4 and 5. }
\label{table:16o}
\end{table}

We simulate the correlated eigenstate of $^{16}$O using the two chosen Skyrme forces, SV and SHZ2. The summary of results shown in Table \ref{table:16o} indicates an insignificant difference between the two interactions, for both values of $N_\text{max}$. For $N_\text{max}=14$, the correlation energy is $-4.1$ MeV for SV and SHZ2. As one increases to $N_\text{max}=20$, the correlation energy increases to $-5.2$ MeV for SV and $-5.1$ MeV for SHZ2. Minute differences are also found between the radii predicted by  the two interactions. 
We ascribe these minutes differences to the fact that the two Skyrme forces, themselves, are very similar to each other. 

\begin{figure}[t]
\centering
\includegraphics[width=\linewidth]{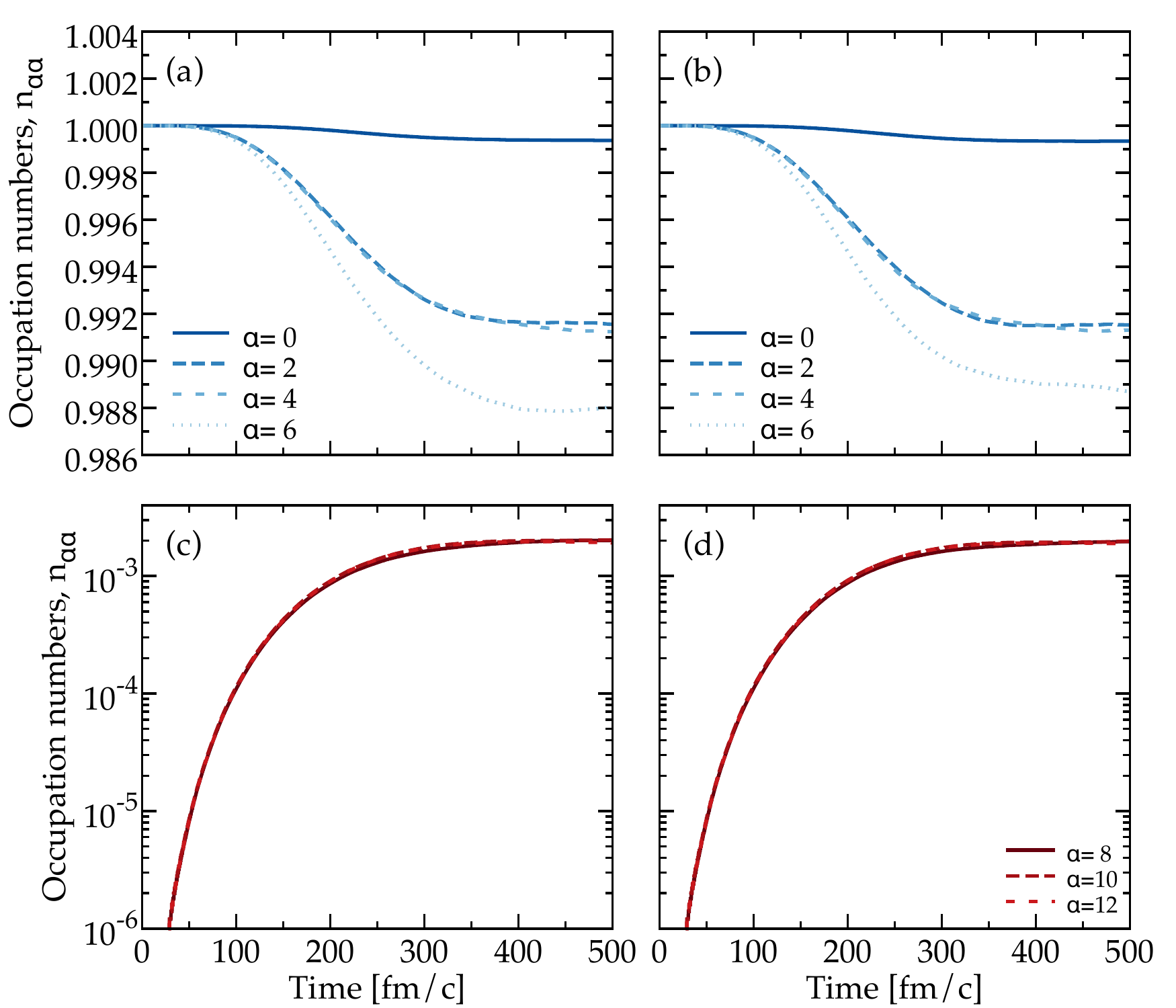}
\caption{Same as Fig.~\ref{fig:n_C12} for $^{16}$O with $N_\text{max}=14$. }
  \label{fig:n_O16}
\end{figure}

As we have already seen, the occupation probabilities $n_{\alpha \alpha}$ provide a way of characterising correlations. Their time evolution within the adiabatic switch-on process for $^{16}$O is shown in Fig.~\ref{fig:n_O16}. These have been obtained with the SV interaction in the $N_\text{max}=14$ model space. For $^{16}$O, HF simulations include $16$ single-particle orbitals corresponding to the $1s_{1/2}$, $1p_{3/2}$ and $1p_{1/2}$ neutron and proton hole subshells.
Like the $^{12}$C case, the deeply bound $1s$ state  (solid line in panels (a) and (b) of Fig.~\ref{fig:n_O16}) remains practically fully occupied. In contrast to the previous case, the occupation of the hole $1p$ states [dashed and dotted lines in panels (a) and (b)] is at level of $\approx 99\%$. This indicates a far less correlated eigenstate than $^{12}C$, where the same orbital was depleted by almost $5\%$. 

The particle states of panels (c) and (d) tell a similar story. Whereas levels close to the Fermi surface for $^{12}$C reached relatively large occupations of order $10^{-2}$, all (degenerate $1d_{5/2}$) particle states of $^{16}$O show a much smaller final occupation, close to $0.002$. We note that the final states are static in terms of the adiabatic switch-on. Just as in the case of $^{12}$C, the diagonal elements of the single-particle energy matrix shown in panel (b) of Fig.~\ref{fig:spes} are relatively constant across the adiabatic evolution for the $N_\text{max}=14$ model space. Unlike the high-lying $^{12}$C particle states, all particle states in $^{16}$O are bound so we do not anticipate any continuum discretization issues in our simulations.

$^{16}$O has been used as a benchmark nuclear system in the past, including different implementations of TDDM~\cite{Tohyama1998b,Tohyama2013,Tohyama2015,Assie2009,AssiePhD}. These studies have generally relied on different mean-field and residual interactions; have neglected spin-orbit couplings in the residual channel and/or have restricted the relevant correlation model space to $p$ and $d$ subshell orbitals. The results typically obtained in these models are much more correlated than those discussed here. Typical $p-$shell ($d$-shell) orbital occupations are closer to $90\%$ ($10\%$), and correlation energies are $\gtrapprox  -10$ MeV. Without a more in-depth analysis and lacking unbiased benchmarks, it is difficult to find a clear explanation for the origin of these discrepancies.

\subsection{\texorpdfstring{$^{20}$Ne}{20Ne}}

\begin{table}[t!] 
\centering
  \begin{tabular}{ | l | l | l | l | l |   }
    \hline
    & \multicolumn{2}{|p{2.1cm}|}{$^{20}$Ne SV} & \multicolumn{2}{|p{2.1cm}|}{$^{20}$Ne SHZ2}  \\ \hline    
    $N_\text{max}$          & $14$ & $20$ &   $14$ & $20$ \\ \hline
    $E_\text{c}$ [MeV]     & $-2.1$ & $-7.3$ & $-2.0$ & $-5.9$ 
    \\ \hline
    $E_\text{MF}$ [MeV] & $-137.7$ & $-134.4$  & $-138.1$ & $-135.6$ 
    \\ \hline
    $E$ [MeV]                 & $-139.8$ & $-141.7$ & $-140.1$ & $-141.5$ 
    \\ \hline
    $E_\text{c}/E$ [\%]  & $1.5$ & $5.15$ & $1.43$ & $4.17$ 
    \\ \hline
   Proton & $2.955$ & $2.980$ &  $2.950$ & $2.969$  \\
     rms [fm] &  &  & &  \\
   \hline
   Neutron & $2.930$ & $2.946$ &  $2.922$ & $2.938$  \\ 
      rms [fm] &  &   &  &   \\ 
    \hline
  \end{tabular}
\caption{The same as Table~\ref{table:12c} for $^{20}$Ne. Results for SHZ2 are also provided. }
\label{table:20Ne}
\end{table}

We discuss the isotope $^{20}$Ne as the first of a series of examples centered around $A=20$. This region of the chart has received significant experimental attention \cite{Schumaker2008,Schumaker2009} due, among other things, to its relevance for astrophysics \cite{dauria2004}. In theoretical studies, this region is typically accessed theoretically using the shell model and is of particular interest in the context of isospin symmetry breaking~\cite{Bentley2007}. In a standard shell model picture, $^{20}$Ne is built from an $^{16}$O core by adding two neutrons and two protons. It is, of course, yet another $n\alpha$ system. 

\begin{figure}[t!]
\centering
\includegraphics[width=\linewidth]{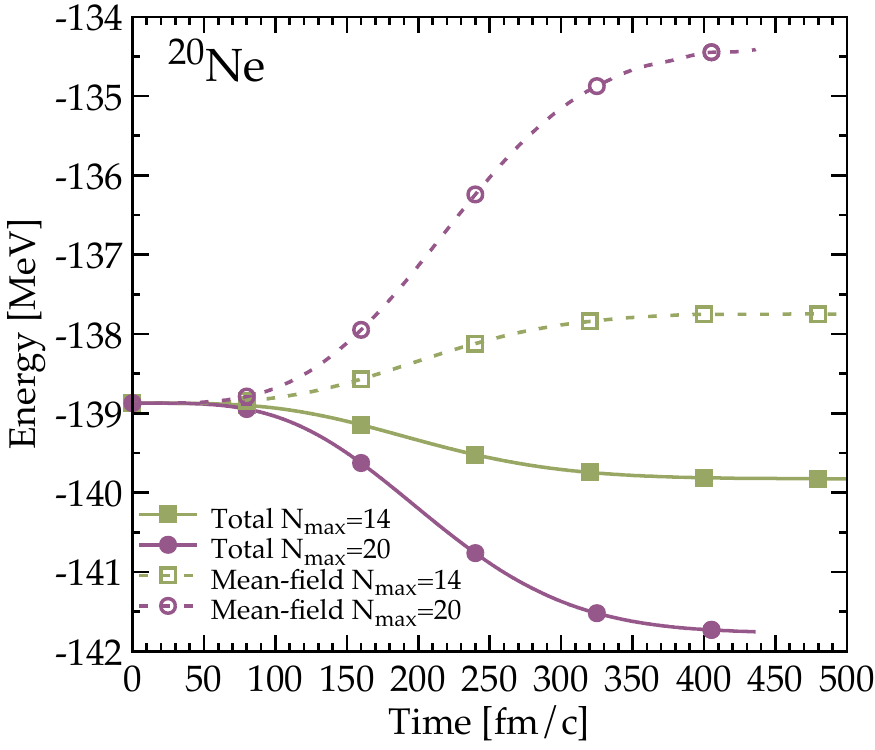}
\caption{Same as Fig.~\ref{fig4:sfig2} for $^{20}$Ne.}
\label{fig4:sfig4}
\end{figure}

We provide a figure for the time evolution of the mean-field (dashed lines) and total (solid lines) energies of $^{20}$Ne in Fig.~\ref{fig4:sfig4}. As with previous cases, the results are shown for two values of $N_\text{max}$ for the SV force. The $N_\text{max}=14$ results (squares) converge well with time, and show no signs of oscillations at late times. The $N_\text{max}=20$ simulation stopped some time before $500$ fm/c, but the results appear to be relatively well converged at this level. A key difference between the results shown in this figure and those of previous isotopes is the relatively large difference in energies between the results obtained with the two model spaces. When going from $N_\text{max}=14$ to $20$, the total energy decreases by almost $2$ MeV - more than double the result observed in other isotopes. Another striking feature is the large increase in the ratio $E_C/E$, which more than doubles when going from one model space to the other. We interpret these results as a sign that the $1d_{5/2}$ subshell closure obtained in the $N_\text{max}=14$  model space is relatively weak in this nucleus. The full $sd$ shell of the $N_\text{max}=20$ simulation provides a much more complete picture that substantially enhances the correlation of the system. 

Table~\ref{table:20Ne} provides a breakdown of the energy contributions for the two Skyrme forces, SV and SHZ2. For $N_\text{max}=14$, the results obtained with these two parameterisations are almost indistinguishable in terms of correlation energy. As one increases the model space to $N_\text{max}=20$, both parametrisations predict the aforementioned substantial increase in correlation energies, from $-2.1$ MeV to $-7.4$ ($-5.9$) MeV for the SV (SHZ2) force. 

\subsection{\texorpdfstring{$^{21}$Ne}{21Ne}}

\begin{table}[t!]
\centering
  \begin{tabular}{ | l | l | l | l | l |  }
    \hline
    \multicolumn{3}{|p{2.1cm}|}{$^{21}$Ne SV}  \\ \hline    
    $N_\text{max}$          & $14$ & $20$ \\ \hline
    $E_\text{c}$ [MeV]     & $-1.3$  &  $-6.1$ 
    \\ \hline
    $E_\text{MF}$ [MeV] & $-146.2$ & $-143.3$ 
    \\ \hline
    $E$ [MeV]       & $-147.5$ & $-149.4$
    \\ \hline
    $E_\text{c}/E$ [\%]  & $0.88$ & $4.08$ 
    \\ \hline
  \end{tabular}
\caption{Energetics of the $^{21}$Ne ground states for different values of $N_\text{max}$ obtained with the SV force.}
\label{table:21Ne}
\end{table}

We turn our attention to an open-shell, odd-even and deformed system, $^{21}$Ne, to confirm that such systems can be tackled with our approach. 

For $^{21}$Ne, Table~\ref{table:21Ne} summarises the energy contributions for the two values of $N_\text{max}$. Just like in the previous case, we find a substantial increase of the correlation energy (more than a factor of $4$) when moving from $N_\text{max}=14$ to $20$. In turn, the relative contribution to the energy increases from below $1\%$ to over $4\%$. This clearly indicates the importance of $sd$ shell contributions for this region of the nuclear chart. 

We also find interesting systematics when comparing the $^{21}$Ne results of Table~\ref{table:21Ne} to the $^{20}$Ne simulations presented in Table~\ref{table:20Ne}. With the addition of one neutron on top of $^{20}$Ne, for instance, we observe that the correlation energy drops by $0.8$ ($1.2$) MeV, in the case of $N_\text{max}=14$ ($20$). This drop in correlation energy can be understood naively, in terms of 
a reduction in the number of neutron levels available to scatter into. As for the total energy, the Hartree-Fock prediction for $^{21}$Ne is $\approx 8$ MeV more bound than $^{20}$Ne. The TDDM ground states energies of the two isotopes differ by $7.7$ MeV, indicating that isotopic differences in the binding energy are largely unchanged by correlations. Interestingly, this occurs because the mean-field contribution to the isotopic difference largely cancels the correlation one. 

Some additional features of this simulation are further reported in Ref.~\cite{BartonPhD}. We note, in particular, that the adiabatic switching-on process for $^{21}$Ne is such that, for both $N_\text{max}=14$ and $20$, the proton and neutron radius did not converge to a static result. This indicates that the transition from the mean-field to the correlated state is more difficult than in some of the previous examples, possibly because of the odd-even nature of the isotope. We also performed an analysis of some of the mean-field energy components, not provided here for brevity. 
The data for the $t_0$ component of the mean-field (which is proportional to the overall density of the system) for $N_\text{max}=14$ shows an increase of 
$\approx 2 $ MeV. The same component for $^{20}$Ne, in contrast, went up by over $3$ MeV. We take this as an indication that the single addition of a neutron can change significantly how different components of the Skyrme force change beyond the mean field limit.   

\subsection{\texorpdfstring{$^{21}$Na}{21Na}}
\begin{table}
\centering
  \begin{tabular}{ | l | l | l | l | l |  }
    \hline
    \multicolumn{3}{|p{2.1cm}|}{$^{21}$Na SV}  \\ \hline    
    $N_\text{max}$          & $14$ & $20$ \\ \hline
    $E_\text{c}$ [MeV]     & $-1.4$ & $-5.9$
    \\ \hline
    $E_\text{MF}$ [MeV] & $-142.2$ & $-139.5$
    \\ \hline
    $E$ [MeV]                 & $-143.6$ & $-145.4$
    \\ \hline
    $E_\text{c}/E$ [\%]  & $0.97$ & $4.05$
    \\ \hline
  \end{tabular}
\caption{The same as Table~\ref{table:21Ne} for $^{21}$Na.}
\label{table:21Na}
\end{table}

We continue our analysis by considering $^{21}$Na, the mirror nucleus to  $^{21}$Ne with an odd proton number. This provides an interesting insight into the nature of isospin symmetry not only at the mean-field, but also at the TDDM level.   
The different energy contributions for $^{21}$Na are provided in Table~\ref{table:21Na} for two values of $N_\text{max}$. 
We find results that bode well with those observed for the isospin partner nucleus. First, as observed for the two previous isotopes, we find that the correlation energy increases substantially with the model space size: form $-1.4$ MeV for $N_\text{max}=14$ to $-5.9$ for $N_\text{max}=20$. This corresponds to almost a factor of $4$ in the relative contribution of the correlation energy, which increases from about $1\%$ to $4 \%$. 
Second, comparing the correlation energy obtained for $^{20}$Ne with that of $^{21}$Na for $N_\text{max}=14$ ($20$) one sees that the addition of one proton reduces the magnitude of the correlation energy by $0.7$ ($1.4$) MeV. This mirrors the reduction we found for $^{21}$Ne compared to $^{20}$Ne. Again, this is presumably due to the reduction in the number of levels available for the nucleons to scatter into. 

The results in tables \ref{table:21Ne} and \ref{table:21Na} allow us to analyse the level of isospin symmetry in our TDDM simulations. At the mean-field level, the results of Table \ref{table:tddmhf} indicate a binding energy difference between the two isotopes of $\approx 3.8$ MeV, close to the experimental difference of $\approx 4.3$ MeV. The origin of this difference can be ascribed mostly to the Coulomb interaction, which is explicitly included in the mean-field simulation using the standard exchange approximation. At the TDDM level, we treat the Coulomb interaction in a cruder way to avoid computing every (finite-range) proton-proton Coulomb interaction matrix element. Instead, every proton-proton interaction matrix element is given an equal proportion of the Coulomb mean-field contribution including the density-dependent exchange term. As a result, we find that the total energy difference between the two isotopes remains very close to $4$ MeV, regardless of the model space, the same value as obtained in the mean-field simulation within our method uncertainties. 

Looking at the specifics, we find that, with $N_\text{max}=20$,$^{21}$Ne produces -$6.1$ of correlation energy, while $^{21}$Na yields -$5.9$ MeV. This is a very small difference of order a few percent, close to the accuracy expected in the extraction of these quantities from our adiabatic switch-on method. In other words, we do not find a significant contribution of correlations to the mass difference of isospin partners. 

\subsection{\texorpdfstring{$^{22}$Na}{22Na}}
\begin{table}
\centering
  \begin{tabular}{ | l | l | }
    \hline
    \multicolumn{2}{|p{2.1cm}|}{$^{22}$Na SV}  \\ \hline    
    $N_\text{max}$  & $20$ \\ \hline
    $E_\text{c}$ [MeV] & $-7.0$ 
    \\ \hline
    $E_\text{MF}$ [MeV] & $-149.1$
    \\ \hline
    $E$ [MeV] &  $-156.1$
    \\ \hline
    $E_\text{c}/E$ [\%] & $4.48$
    \\ \hline
  \end{tabular}
\caption{The same as Table~\ref{table:21Ne} for $^{22}$Na, but for a single value of $N_\text{max}$. }
\label{table:22Na}
\end{table}

To finish the discussion in the $A=20-22$ mass region, we discuss the proof-of-principle case of $^{22}$Na, a nucleus with an odd number of neutrons and protons.
We encountered some technical issues in attempting to run  simulations of this isotope with $N_\text{max}=14$. We found, for instance, energy level crossings, which precluded us from identifying the final state of the evolution with the ground state of the system.  
In addition, as the single-particle energies crossed, the occupations of both levels approached $\approx 0.5$, indicating a strong departure from the single-particle picture. Finally, the time evolution of the energy departed significantly from the expected $\gamma^2(t)$ dependence associated to the Born term. All in all, the $N_\text{max}=14$ results indicate that correlations are significantly changing the  structure of this nucleus. It is possible that an insufficiently large model space cannot capture this significant changes in the switching procedure. 

In contrast, the numerics for the $N_\text{max}=20$ case were remarkably more stable. 
Table~\ref{table:22Na} provides a summary of the energetics obtained for this isotope, focusing only on the $N_\text{max}=20$ results. We find a relatively large correlation energy of $7$ MeV, which is about $\approx 1$ MeV larger than the neighbouring $A=21$ isotopes and in good agreement with the $^{20}$Ne result. In relative terms, this is about $\approx 4.5 \%$ of the total energy, close to the value that we have observed across this mass region. In other words, it appears that the instability in the $N_\text{max}=14$ results does not significantly reflect in the converged results.

\subsection{\texorpdfstring{$^{24}$O}{24O}}
\begin{table}
\centering
  \begin{tabular}{ | l | l |   }
    \hline
    \multicolumn{2}{|p{2.1cm}|}{$^{24}$O SV}  \\ \hline    
    $N_\text{max}$         & $20$ \\ \hline
    $E_\text{c}$ [MeV]     &$-4.6$  
    \\ \hline
    $E_\text{MF}$ [MeV] & $-142.0$ 
    \\ \hline
    $E$ [MeV]                 & $-146.6$ 
    \\ \hline
    $E_\text{c}/E$ [\%]  & $3.14$ 
    \\ \hline
  \end{tabular}
\caption{The same as Table~\ref{table:22Na} for  $^{24}$O.}
\label{table:24O}
\end{table}

We finish our discussion with an exotic, neutron-rich isotope: $^{24}$O. This provides a test case for a nucleus relatively far from stability with a significant asymmetry between proton and neutrons. This isotope is indeed close to the neutron drip line and is at the center of a series of contemporary experimental developments  \cite{PhysRevLett.102.152501,PhysRevLett.100.152502,PhysRevC.92.051306,PhysRevLett.109.022501}. 
Importantly, some results for this isotope have been previously reported in other TDDM implementations \cite{Tohyama2002b,Assie2009,AssiePhD}. 
Our calculations were performed with the SV parameterisation and a model space with $N_\text{max}=20$. The results are summarised in Table \ref{table:24O}. 

We predict a correlation energy in $^{24}$O which is $\approx 4.6$ MeV. This value is obtained by extrapolating data in the time evolution up to $\approx 250$ fm/c.
Tohyama and Umar report a correlation energy of $-3.5$ MeV for $^{24}$O in Ref.~\cite{Tohyama2002b}, whereas Assi\'e and Lacroix find $-4.6$ MeV using the TDDM$^P$ approach. Both values bode relatively well with our finding, even though they have been obtained with different mean-field (and residual) interactions. More importantly, these predictions rely on using only a handful of active orbitals and, typically, an $^{16}$O inert core. 

Compared to the equivalent results for the symmetric isotope  $^{16}$O in Table~\ref{table:16o}, the correlation energy has decreased by about $0.5$ MeV when increasing the neutron number from $N=8$ to $16$. This follows the qualitative trend discussed in previous isotopes, which indicates a reduction of correlation energy as neutron number increases. These findings bode well with the idea that, in increasing neutron number, there are less orbitals to scatter into and, hence, less of a correlation energy. 
This decrease  is also consistent with the isotopic evolution reported in oxygen both in Refs.~\cite{Tohyama2002b} and \cite{AssiePhD}. 
The results shown in Ref.~\cite{Tohyama2002b} when going from $^{22}$O to $^{24}$O show a decrease of almost $1$ MeV. The  no-core simulations in Ref.~\cite{AssiePhD} also show a decrease of $E_c$ with neutron number, although the order of magnitude of the correlation is different. 

Our simulations are also influenced by the closeness to the drip line. In the HF simulation, all $16$ neutron states are bound. Upon switching correlations on with TDDM, however, some of the unbound HF states become occupied through beyond mean-field scattering. In particular, there are $4$ levels that are very close to being bound with energies $\approx 0.25$ MeV. These almost-bound levels also have a relative occupation which is two orders of magnitude larger than any of the remaining unbound states. In a naive shell model interpretation, one would interpret these $4$ additional as those filling the neutron $sd$ shell. The treatment of a box-discretized continuum may be somewhat inadequate here, but it does not preclude the convergence of our simulations.
While we have not explored these effects further, it is possible that adding more neutrons to oxygen isotopes may shift occupations and single-particle energies in a way that the mean-field simulation cannot capture, providing perhaps a different neutron drip line in TDDM than in HF. 

\subsection{Correlation entropy}

\begin{figure}
\centering
\includegraphics[width=\linewidth]{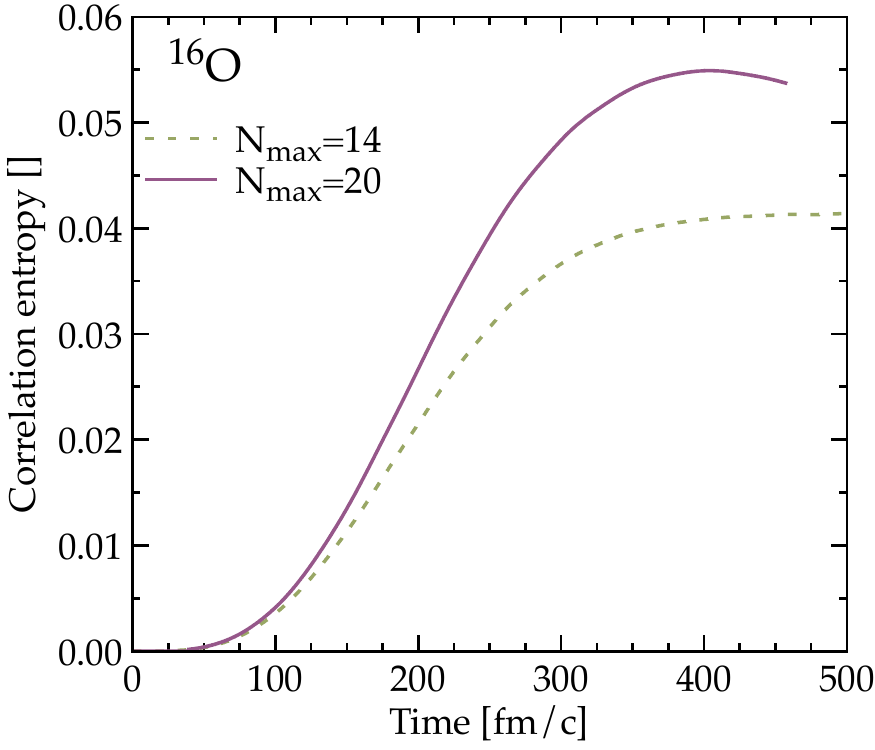}
\caption{Correlation entropy of $^{16}$O as a function of time for two different model spaces with $N_\text{max}=14$ (dashed line) and $20$ (solid line). }
\label{fig:entropy}
\end{figure}

Up to this point, we have looked at the effect of correlations on specific, measurable single-particle and bulk nuclear properties. There are however other theoretical measures of correlations that provide independent characterisations. One of such measures is the so-called correlation entropy \cite{entropy,entropy2}, which is computed from the diagonal occupation numbers $n_{\alpha \alpha}$ as
\begin{equation}
S_\text{cor}  = \ - \dfrac{1}{A} \sum_{\alpha} n_{\alpha \alpha} \ln n_{\alpha \alpha} \, ,
\end{equation}
where A is the number of nucleons.
This quantity is, strictly speaking, not an entropy from a thermodynamic point of view, but rather an approximation of one as, among other things, one neglects the off-diagonal occupation matrix elements \cite{entropy2}. 
The correlation entropy is exactly $0$ for the HF state, and necessarily increases as one goes towards a correlated eigenstate.  The absolute numerical quantity does not have a direct physical meaning, but comparison of values between different calculations can be instructive. Naively, one expects more ``correlated" ground states to yield larger values for $S_\text{cor}$, in the sense that they depart more from the HF eigenstates of $0$ entropy.

A graph of this quantity as a single $^{16}$O nucleus goes from the HF eigenstate to the correlated eigenstate, for both $N_\text{max}=14$ (dashed line) and $20$ (solid), is shown in Fig.~\ref{fig:entropy}. As expected, in the initial HF state both simulations yield $0$ correlation entropy. As the system evolves towards the correlated ground state, the entropy steadily increases until it levels off around $300$ fm/c into the evolution. 
We note that the entropy does not completely converge at the end of the calculation, particularly for $N_\text{max}=20$, where the simulations suggest the occurrence of a maximum of $S_\text{cor}$ at intermediate times. We take this as an indication that the occupation numbers $n_{\alpha \alpha}$ are not entirely converged, which in turn suggests that the adiabatic switching time is relatively small. Neither the energy nor the occupation numbers reported in Figs.~\ref{fig4:sfig3} and \ref{fig:n_O16}, however, showed a clear non-stationarity at the end of the simulation. 

Figure~\ref{fig:entropy} also suggests that the correlation entropy increases with $N_\text{max}$. 
We find that this is a generic feature at least in the the two model spaces explored here. 
Table \ref{entropytable} shows the correlation entropy obtained numerically at the end of the adiabatic evolution for the various nuclei studied in this work. 
In all cases, the entropy computed with $N_\text{max}=14$ states is smaller than that computed with $N_\text{max}=20$. In a sense, this can be understood naively, in that additional levels necessarily provide more contributions to the correlation entropy. In this sense, the correlation energy is not a good measure of the model-space convergence of the results. 

Two more features stand out from the results on Table~\ref{entropytable}. First, we find that the nuclei with  large correlation energies, like $^{12}$C, also have large correlation entropies. Second, we find that the relative increase in correlation entropy when going from $N_\text{max}=14$ to $20$ is very similar to the corresponding relative increase in correlation energies. In other words, we find that both the correlation energy and the correlation entropy provide relatively similar measures of correlations in the systems that we have studied. We note that this is not necessarily trivial \emph{a priori}. The calculation of the correlation entropy relies entirely on one-body occupation numbers, whereas the correlation energy is the result of the contraction of 2 seemingly different two-body objects - the interaction and the correlation tensors, see Eq.~(\ref{eq:ecor}). 

\begin{table}[t] 
\centering
  \begin{tabular}{ | l | l | l | }
    \hline
     \multicolumn{3}{|p{2.1cm}|}{SV}  \\ \hline
		    $N_\text{max}$ & $14$ & $20$
    \\ \hline
        $^{12}$C & 0.088  & 0.132 \\
    \hline
   		$^{16}$O & 0.041 & 0.053 \\
    \hline
        $^{20}$Ne & 0.020 & 0.073 \\
    \hline
        $^{21}$Ne & 0.010 & 0.041 \\
    \hline
        $^{21}$Na & 0.010 & 0.038 \\
    \hline
    \multicolumn{3}{|p{2.1cm}|}{SHZ2}  \\ \hline
		    $N_\text{max}$ & $14$ & $20$
    \\ \hline
       		$^{16}$O & 0.044 & 0.055 \\
    \hline
       		$^{20}$Ne & 0.020 & 0.065 \\
    \hline
  \end{tabular}
\caption{Correlation entropies in the TDDM ground state of all the isotopes considered in our work for two values of $N_\text{max}$ and two Skyrme parametrizations.}
\label{entropytable}
\end{table}

\section{Conclusions and future outlook}
\label{sec:conclusions}

In this paper, we implemented the TDDM approach including up to two-body correlations to study nuclear ground states. Unlike some of the previous work in the field, our simulations are performed without any spatial symmetry restrictions, following well-established precursors in TDHF \cite{sky3dcode}. We also use a self-consistent interaction, both at the mean-field level and at the residual one. To this end, we work with density-independent Skyrme interactions to avoid any issues related to rearrangement terms. We compute the ground states using a dynamical TDDM code. Our starting point is the corresponding HF ground state. We then switch-on beyond mean-field correlations adiabatically, by slowly incorporating  beyond-mean-field terms over time.

With this approach, we investigate light nuclei with masses ranging from $A=12$ to $A=24$. Our approach can tackle closed-shell nuclei, like $^{12}$C and $^{16}$O, but also open-shell isotopes, like $^{20-21}$Ne, $^{21-22}$Na or $^{24}$O. We find correlated ground states for all these isotopes. We study the effect of TDDM correlations using a variety of quantities, including single-particle energies and occupations, but the main focus of our analysis is on binding energies. For the majority of nuclei, the correlation energy accounts for $\approx 4$ to $5 \%$ of the total energy. A clear exception to this trend is $^{12}$C, where two-body correlations are significantly stronger and account for $\approx 11\%$ of the total energy. A quantitative metric based on the correlation entropy provides similar results. Where the comparisons are possible, our results provide qualitatively similar predictions to previous TDDM implementations. We also confirm a trend of diminishing correlations when the neutron number increases. 

We find two key limitations in this initial study, that could be improved in the future. On the one hand, the Skyrme parametrizations that we have used are relatively poor. Among other things, they have not been fitted to this mass region or to account for beyond mean-field correlations and, in this sense, our predictions can only indicate qualitatively the size of correlations in these nuclei. On the other hand, we find that our adiabatic switching produces final states that may appear static when it comes to one observable, like the energy, but are not stationary in others, like the correlation entropy. This indicates that longer evolution times are required, although this comes at a significant larger numerical cost. 

This work opens several potential avenues for immediate future work. When it comes to computing ground state properties, we have demonstrated that TDDM provides a stable description of relatively light nuclear systems. The extension to higher mass numbers is straightforward, if numerically challenging. One could further characterise these TDDM ground-state by analysing their cluster structure or by exploiting the connections between TDDM and different pairing approximations. 
Furthermore, time-dependent techniques are particularly suitable for the study of excitations on top of these ground states. It may be interesting to excite and time-evolve different modes using TDDM, to test the validity of mean-field approaches but also to identify correlation effects on resonances. Finally, dynamical simulations can also tackle nuclear collisions of interest for a variety of application, including heavy-ion fusion reactions~\cite{Wen2018}. 

We can also envisage some additional formal developments that could be useful in the context of nuclear physics. One could attempt to overcome the limitations associated to density-independent forces by extending the TDDM formalism to include genuine three-nucleon interactions. This is presumably challenging, since the BBGKY hierarchy would likely have to be modified. The treatment of genuine three-body correlations may be relevant in nuclear systems too~\cite{Tohyama2020}. To break away from the adiabatic evolution picture, one could also attempt to devise an energy minimisation process that included two-body density matrices~\cite{garrod2,garrod,var2,var1}. 
By implementing beyond mean-field simulations, like those presented here, together with the aforementioned developments, one would open the door to a truly first-principles understanding of these applications.

\begin{acknowledgments}
This work was supported by the Polish National Science Centre (NCN) under Contract No. UMO-2016/23/B/ST2/01789;
by STFC, through Grants Nos ST/M503824/1, ST/L005743/1 and ST/P005314/1; 
and by the Spanish State Agency for Research of the Spanish Ministry of Science and Innovation through 
the ``Ram\'on y Cajal" program with grant RYC2018-026072 and
the ``Unit of Excellence Mar\'ia de Maeztu 2020-2023" award to the Institute of Cosmos Sciences (CEX2019-000918-M).
This work used the DiRAC Complexity system, operated by the University of Leicester IT Services, which forms part of the STFC DiRAC HPC Facility. 
This equipment is funded by BIS National E-Infrastructure capital grant ST/K000373/1 and  STFC DiRAC Operations grant ST/K0003259/1. DiRAC is part of the National E-Infrastructure. 
\end{acknowledgments}

\appendix
\section{\texorpdfstring{$4^\text{th}$ order Runge-Kutta timestep implementation}{4th order Runge-Kutta timestep implementation}}

In typical implementations of TDHF, the time-stepping procedure involves an integration via the midpoint method. We have found that the solution of the TDDM equations necessarily requires time stepping algorithms that provide more accurate results for values of $dt$, the time-step size, that are not prohibitively small. We have therefore implemented an explicit $4^\text{th}$ order Runge-Kutta (RK4) algorithm to solve the set of coupled differential equations of relevance. 
We note that RK4 carries a cumulative error of order $dt^{4}$ \cite{Suli2003}. 

In the case of TDDM, one has a set of three differential equations for the evolution of the single particle orbitals, Eq.~(\ref{eq:spTDHF}); occupations, Eq.~(\ref{eq:occmat}); and correlation tensor elements, [Eq.~(\ref{corten}). We can schematically write this set of equations as follows:
\begin{align}
\dfrac{d \psi}{dt} &=\mathcal{P}(t,\psi,n,C)  \, , \\
\dfrac{d n}{dt} &= \mathcal{N}(t, \psi,n,C)  \, , \\
\dfrac{d C}{dt} &= \mathcal{C} (t,\psi,n,C)    \, .
\end{align}
Using a RK4 algorithm, given the initial conditions $(t_{p},\psi_{p},n_p, C_p)$,  the estimates for the functions at $t_{p+1}$ read:
\begin{align}
\psi_{p+1} &=  \psi_p + \dfrac{dt}{6} \Bigg[ d_{1} + 2 d_{2} + 2 d_{3} + d_{4} \Bigg] \, , \\
n _{p+1}  &= n _p + \dfrac{dt}{6} \Bigg[ e_{1} + 2 e_{2} + 2 e_{3} + e_{4} \Bigg] \, , \\
C _{p+1} &= C _p + \dfrac{dt}{6} \Bigg[ f_{1} + 2 f_{2} + 2 f_{3} + f_{4} \Bigg] \, .
\end{align}
The coefficients $d_{1} \cdots d_{4} $ are given by the following $4$ equations evaluated either at the initial step, at the midpoints or at the final one:
\begin{align} 
d_{1} &= \mathcal{P} \left( t_{p},\psi _p,n_p,C_p \right)  \\
d_{2} &= \mathcal{P} \left( t_{p}+\dfrac{dt}{2},\psi _p+\dfrac{dt}{2}d_{1},n_p+\dfrac{dt}{2}e_{1},C_p+\dfrac{dt}{2}f_{1} \right)  \\
d_{3} &= \mathcal{P} \left( t_{p}+\dfrac{dt}{2},\psi _p+\dfrac{dt}{2}d_{2},n_p+\dfrac{dt}{2}e_{2},C_p+\dfrac{dt}{2}f_{2} \right) \\
d_{4} &= \mathcal{P} \left( t_{p}+dt,\psi _p+dt \ d_{3},n_p+dt \ e_{3},C_p+dt \ f_{3} \right)  .
\end{align}
The remaining coefficients
$e_{1} \cdots e_{4}$ and $f_{1} \cdots f_{4}$ are found analogously using the replacements $\mathcal{P} \to \mathcal{N}$ and $\mathcal{P} \to \mathcal{C}$, respectively. 

We note that, in addition to the single particle orbitals, occupation matrices and correlation tensors, other auxiliary quantities, such as densities and mean-fields, need to be recalculated in the $4$ steps involving $d_i$, $e_i$ and $f_i$. This guarantees the stability of the RK4 algorithm within the TDDM method. In all the simulations performed in this work, we found that $dt \leq 0.2$ fm c$^{-1}$ provided acceptable and numerically stable results. 

\bibliographystyle{apsrev4-1}
\bibliography{Thesisbibliography}

\end{document}